\documentclass{article}
\usepackage{amsmath}
\usepackage{multirow}
\usepackage{booktabs}
\usepackage{cite}
\usepackage{graphicx}
\usepackage{subfigure}
\usepackage[width=.84\textwidth]{caption}

\usepackage[T1]{fontenc}
\usepackage[utf8]{inputenc}
\usepackage{authblk}

\title{Model selection criteria of the standard censored regression model based on the bootstrap sample augmentation mechanism}
\author[1]{Yue Su\thanks{Corresponding author:yuesueactuary@mail.dlut.edu.cn}}
\author[2]{Patrick Kandege Mwanakatwe}
\affil[1]{School of Mathematical Science, Dalian University of Technology}
\affil[2]{School of Mathematical Science, Dalian University of Technology}

\begin{document}
\maketitle

\abstract{
The statistical regression technique is an extraordinarily essential data fitting tool to explore the potential possible generation mechanism of the random phenomenon. Therefore, the model selection or the variable selection is becoming extremely important so as to identify the most appropriate model with the most optimal explanation effect on the interesting response. In this paper, we discuss and compare the bootstrap-based model selection criteria on the standard censored regression model (Tobit regression model) under the circumstance of limited observation information. The Monte Carlo numerical evidence demonstrates that the performances of the model selection criteria based on the bootstrap sample augmentation strategy will become more competitive than their alternative ones, such as the Akaike Information Criterion (AIC) and the Bayesian Information Criterion (BIC) etc. under the circumstance of the inadequate observation information. Meanwhile, the numerical simulation experiments further demonstrate that the model identification risk due to the deficiency of the data information, such as the high censoring rate and rather limited number of observations, can be adequately compensated by increasing the scientific computation cost in terms of the bootstrap sample augmentation strategies. We also apply the recommended bootstrap-based model selection criterion on the Tobit regression model to fit the real fidelity dataset.
}

\section{Introduction}
In the practical data analysis area, regression is an extraordinarily essential data modelling technique used to fit the interesting response variable
so as to capture and identify the potential possible true relationship between the dependent variable and the explanatory variables as accurately as possible.
Therefore, the model selection or variable selection technique is becoming extremely important.

Model selection is based on the Kullback-Liebler (K-L) discrepancy
(Kullback 1951)\cite{kullback1951information} which is a discrepancy measurement
between two different probability distributions.
The Akaike information criterion (Akaike 1973)\cite{akaike1973information}, which is based on the
K-L discrepancy, is a commonly used model selection criterion
used to measure the discrepancy degree between the assumed true model and the corresponding candidate one. By the Taylor expansion and the asymptotical normality of the maximum likelihood
estimation, Akaike showed that the maximized log-likelihood of the model is a
positive biased estimation of the expected log-likelihood and the bias can be
asymptotically approximated by the dimension of the model parameter space. However,
the AIC criterion is not consistent in the sense that the probability of the correct identification on the
true model does not asymptotically tend to one, more specifically, it will asymptotically overshoot the true
model order or the dimension of the parameter space of the true model.
The model selection criteria with the consistency property, such as the Bayesian information criterion (BIC) (Schwarz1978) \cite{schwarz1978estimating} and the
Hannan-Quinn information criterion (HQ) (Hhannan1979)\cite{hannan1979determination}, were consecutively proposed.
The corrected AIC (AICc) (Sugiura1978) \cite{sugiura1978further} was proposed
to improve the model selection performance on the linear regression model under the circumstance of the finite number of observations.
Hurvich and Tsai (Hurvich 1989) \cite{hurvich1989regression} explored the application of the AICc on the nonlinear model and
the auto regression model. The advantage of the AIC criterion is that it can be applied to any model, however,
the derivation of the AICc criterion is highly model related.

The consistency property of the model selection criterion is a statistical large sample property, however the statistical inference efficiency of the model selection
will be jeopardized when the observation information is becoming limited due to some unexpected restrictions, for example it is time or money consuming to acquire
the sufficient observation samples or the acquirement of the sample information is related to some sensitive social and psychological issues.
Meanwhile, the derivation of the model selection criterion is highly related to
the restricted assumptions. For example, the derivation of the AIC information criterion assumes that the potential true probability model
is in the given candidate model class and the data volatility effect is evaluated by the normality property of the maximum likelihood estimate.

To circumvent the analytical difficulty and restricted assumptions and improve the
model identification efficiency especially when the observation information is becoming limited,
bootstrap-based model selection criteria have been proposed amid to use bootstrap methodology to simulate the data fluctuation.
Efron (1979) \cite{efron1979bootstrap} initially introduced the bootstrap methodology as a generalization of the Jackknife and discussed its
advantage when it is used to estimate the bias or variance of the estimator.
Efron (1983) \cite{EfronEstimating} discussed the bootstrap estimation of the error rate of a prediction rule.
Efron and Tibshirani (1986) \cite{efron1986bootstrap} discussed the statistical accuracy of the bootstrap methods.
Ishiguro (1991) \cite{ishiguro1991wic} introduced the bootstrapped model selection criterion known as the WIC (An estimator-free information criterion).
The extension criterion of AIC, known as the EIC, was introduced by Ishiguro (1997)\cite{ishiguro1997bootstrapping}.
Shibata (1997) \cite{shibata1997bootstrap} discussed the bootstrap estimate of Kullback-Leibler information for model selection.
Efron and Tibshirani (1997) \cite{efron1997improvements} introduced the $.632+$ bootstrap method which is an improvements on
the cross-validation. Following Efron and Tibshirani (1997),
Pan (1999) \cite{pan1999bootstrapping} introduced the $.632+$ rule.

The bootstrap-based extensions of the AIC model selection criterion have been applied to different kinds of
models. For example, the bootstrapped variant model selection criterion of AIC for the state space model was discussed by Cavanaugh (1997) \cite{cavanaugh1997bootstrap}.
Bootstrap-based model selection for the mixed model was introduced by Shang (2008) \cite{shang2008bootstrap}. The asymptotic
bootstrap bias for the linear regression model was introduced by Seghouane (2010) \cite{seghouane2010asymptotic}.
Bootstrap-based model selection criterion on the beta regression model was discussed by Bayer (2015) \cite{bayer2015bootstrap}.

In this paper, we mainly discuss and compare the model selection performance of the bootstrap-based model selection criteria on the Tobit regression model
under the circumstances of different bootstrap sample augmentation mechanisms.
Different kinds of bootstrap-based model selection criteria and the sample augmentation strategies on the Tobit regression model are compared by
the Monte Carlo numerical simulation technique. Some useful and empirical recommendations are given based on results of the Monte Carlo simulation experiments.
The recommended bootstrapped model selection criteria based on the Tobit regression model are also applied to fit the real fidelity data.

\section{Motivating dataset example}
Modelling the censoring interesting variable is extremely common in the practical data analysis area.
Tobin (1958) \cite{tobin1958estimation} discussed the estimation of relationships for limited dependent variables in the economic surveys of households.
Amemiya (1973) \cite{amemiya1973regression} considered the parameter estimation of the regression model when the dependent variable is truncated normal.

In this study, the Affairs dataset which is available in the AER package of the
statistical analysis software R will be used to demonstrate our main motivation.
The Affairs dataset consists of number of nine variables and the total number of $601$ observation samples.
The fist variable named Affairs with the relative high censoring rate $0.75$ will be taken as the interesting response variable.
The kernal density curve of the interesting response variable Affairs is demonstrated by the following figure named Kernal Density of Affiars.
The variable Affairs follows the non-Gaussian distribution with a relative high censoring rate and it is appropriate to fit the Affairs variable
by the standard censored regression model.

\begin{figure}[htbp]
\begin{center}
\includegraphics[width=6cm]{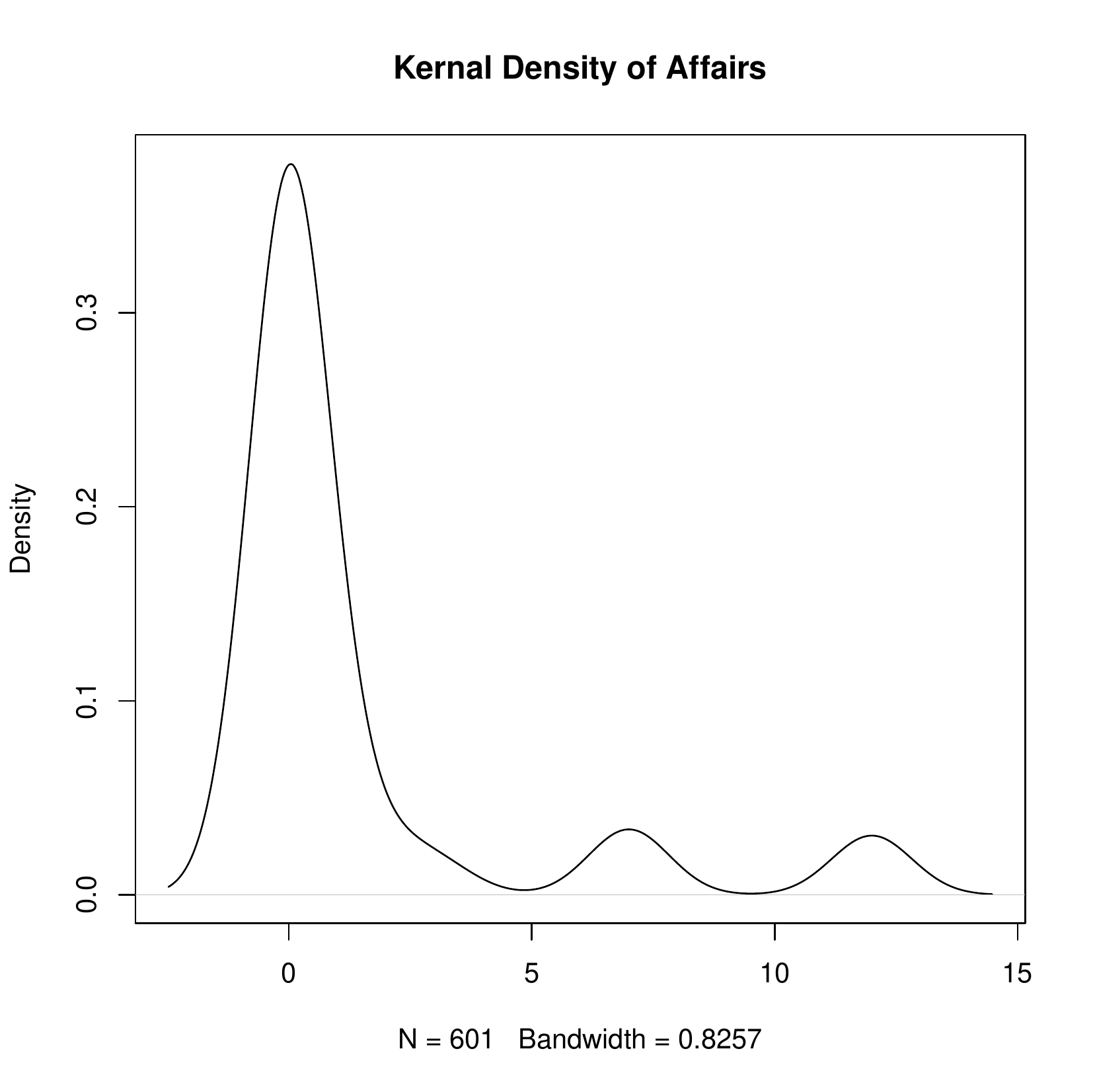}
\end{center}
\end{figure}

The acquirement of the interesting response variable Affairs is time and money consuming because
it is highly related to the social moral constraint and the self-protection of the individual privacy.
Therefore it is indispensable to investigate the model selection performance under the circumstance of the limited observation information.
Bootstrap-based sample augmentation mechanism come into the sight to simulate the potential true data volatility.
To sufficiently demonstrate the model selection performances of different kinds of model selection criteria on the Tobit regression model,
such a total number of $601$ observation individuals in the fidelity data set will be considered as the potential true population with the unknown relationship between the interesting response variable named Affairs and other potential possible explanatory variables.
The objective of the statistical inference is to select the most appropriate candidate model as the optimal relationship expression
between the interesting response and its potential possible explanatory variables based on the limited observation information.
Therefore, increasing the cost of the scientific computation aimed to recuperate the loss of the model identification efficiency brought by the limited
data information constitutes the core issue of the bootstrap-based model selection on the Tobit regression model.

\section{Model selection and its criteria}
\subsection{Model Selection}
Both the statisticians and the practitioners are interested in fitting the random observations by using the statistical model as an approximation expression toward
the potential true generation mechanism of the random phenomenon. If we use the capital letter $Y$ and the notation $g(y)$ to denote
the interesting variable and the corresponding potential true distribution law respectively, the objective of the
model selection is to identify the optimal model from the candidate model class to approximate the potential true distribution law $g(y)$ as accurately as possible.

In order to simplify our discussion and the notation expression, we will not take the different candidate model family with the same dimension of the parameter space
under our consideration. Suppose that we have the specific candidate model class $\mathcal{F}=\left\{\mathcal{F}(1),...,\mathcal{F}(m)\right\}$,
where $m$ is the maximum allowable dimension of the model parameter space $\Theta$.
The candidate model families $\mathcal{F}(k),k=1,...,m$ are sequentially nested in the sense that they satisfy the relationship $\mathcal{F}(1)\subset\mathcal{F}(2)\subset\cdot\cdot\cdot\subset\mathcal{F}(m-1)\subset\mathcal{F}(m)$.
The parsimonious candidate model family $\mathcal{F}(k), k=1,...,m$ is obtained by setting
the number of $m-k$ parameters to constants. Without loss of generality, these constants can be assumed to be zeros. Then the
candidate model family $\mathcal{F}(k)$ can be expressed through the following expression,
\begin{equation*}
\mathcal{F}(k)=\left\{f(y;\theta_{k}):\theta_{k}\in\Theta_{k}\right\}, \Theta_{k}=\left\{(\theta_{1},\theta_{2})^{T}:\theta_{2}=0\in\Theta_{m-k}\right\},
\end{equation*}
where $f(y;\theta_{k})$ is the parametric probability model with the $k$-dimensional model parameter defined in
a $k$-dimensional parameter space $\Theta_{k}$. Furthermore, we use the notation
$\hat{\theta}_{k}$ to express the maximum likelihood estimation of the model with number of $k$ model parameters
and $\hat{\theta}_{k}$ is the solution of the following optimum problem,

\begin{equation*}
\hat{\theta}_{k}=\mathop{\arg\max_{\theta\in\Theta_{k}}} L(\theta),\Theta_{k}=\left\{(\theta_{1},\theta_{2})^{T}: \theta_{2}=0\in \Theta_{m-k}\right\},
\end{equation*}
where $L(\theta)$ is the likelihood function of the parsimonious model with the number of $k$ model parameters.

If we assume that there exists a probabilistic model $f(y;\theta_{k_{0}})\in\mathcal{F}(k_{0})\in\mathcal{F}$ such that $f(y;\theta_{k_{0}})$ is equivalent to the potential true random generation mechanism $g(y)$, the final estimated model $f(y;\hat{\theta}_{k})$ is said to be correctly specified if $f(y;\theta_{k_{0}})\in \mathcal{F}(k)$, but
there does not exist any candidate model family $\mathcal{F}(s)$ such that $f(y;\theta_{k_{0}})\in\mathcal{F}(s)$, where $s<k$.
The final selected model $f(y;\hat{\theta}_{k})$ is said to be over specified if $f(y;\theta_{k_{0}})\in\mathcal{F}(k)$, but
there exists a candidate model family $\mathcal{F}(s)$ such that $f(y;\theta_{k_{0}})\in\mathcal{F}(s)$, where $s<k$.
The final selected estimated model $f(y;\hat{\theta}_{k})$ is said to be under specified if $f(y;\theta_{k_{0}})\notin \mathcal{F}(k)$.
Meanwhile, if the candidate model class $\mathcal{F}$ does not contain any candidate model which can be considered as an equivalent
expression to the potential true distribution law $g(y)$,
the final selected model $f(y;\hat{\theta}_{k})$ will be considered as the optimal approximated model expression toward the potential true random generation mechanism $g(y)$.

\subsection{Model selection criterion}
Model selection is based on the concept of the K-L discrepancy. The candidate model with the minimum K-L discrepancy with
respect to the potential true probability distribution law will be considered
as the optimal fitted model. In this section, we briefly introduce different kinds of model selection criteria which are commonly applied
in the research and practical data analysis realms.

In order to distinguish the meanings of the different notations,
we use the capital letter $Y=(Y_{1},...,Y_{n})^{T}$ to denote the number of $n$ random variables coming from the potential true unknown population and the small letter $y=(y_{1},...,y_{n})^{T}$ to express the corresponding realization. Similarly, we use the capital letter $Y^{b}=(Y_{1}^{b},...,Y_{n}^{b})^{T}$ to express the number of $n$ random variables under the specific bootstrapped generation mechanism and the small letter $y^{b}=(y_{1}^{b},...,y_{n}^{b})^{T}$ to denote its realization.
The notation $\hat{\theta}^{b}_{k}$ stands for the maximum likelihood estimation of the $k$-dimensional model parameter based on the number of $n$ bootstrap observation
$Y^{b}$ and the notation $\hat{\theta}_{k}$ denotes the maximum likelihood estimation of the $k$-dimensional model parameter based on the number of $n$ validated observations of $Y$. We use the notations $E_{Y^{b}}$ and $E_{Y}$ to express taking the expectation with respect to $Y^{b}=(Y_{1}^{b},...,Y_{n}^{b})^{T}$ and $Y=(Y_{1},...,Y_{n})^{T}$ respectively.

\subsubsection{Akaike information criterion (AIC) and its alternatives}
The derivation of the AIC is based on the Kullback Leibler (K-L) divergence
(Kullback 1951) which is a
distance measurement between two different probability distributions.
If we assume that the potential true generation mechanism of the random phenomenon can be
described by the parametric probability model $f(y;\theta_{k_{0}})$,
the K-L divergence between the true model $f(y;\theta_{k_{0}})$ and the
candidate model $f(y;\theta_{k})$ can be described by the following expression,
\begin{equation*}
E_{Y}\left\{\log \left[\frac{f(y;\theta_{k_{0}})}{f(y;\theta_{k})}\right]\right\}, \tag{3.2.1.1}
\end{equation*}
where $\theta_{k_{0}}$ is the $k_{0}$-dimension true model parameter vector and
$\theta_{k}$ is the optimal $k$-dimension model parameter vector in the sense that $\hat{\theta}_{k}$ is the consistent estimation of $\theta_{k}$, where
$\hat{\theta}_{k}$ is the maximum likelihood estimation of the model with the number of $k$ parameters and the number of $n$ observations.

The expression $(3.2.1.1)$ can be further expanded based on the linearity property of the expectation operation as following
\begin{equation*}
E_{Y}\left[\log f(y;\theta_{k_{0}})\right]-E_{Y}\left[\log f(y;\theta_{k})\right]. \tag{3.2.1.2}
\end{equation*}
The first term of the expression $(3.2.1.2)$ $E_{Y}\left[\log f(y;\theta_{k_{0}})\right]$
is completely determined by the potential true model $f(y;\theta_{k_{0}})$.
Therefore, the discrepancy measurement
$(3.2.1.1)$ will be completely determined by the second term of the expression $(3.2.1.2)$ which is the negative of the cross-entropy expression
$E_{Y}\left[\log f(y;\theta_{k})\right]$.
Nevertheless, it is impossible to accurately quantify the cross-entropy $E_{Y}\left[\log f(y;\theta_{k})\right]$ because the
true probability model $f(y;\theta_{k_{0}})$ is unknown. Akaike (1973) recommended to use the maximized log-likelihood expression
$\log f(y;\hat{\theta}_{k})$ as an estimation on the cross-entropy $E_{Y}\left[\log f(y;\hat{\theta}_{k})\right]$ and
showed that the following bias expression

\begin{equation*}
E_{Y}\left\{\log f(y;\hat{\theta}_{k})-E_{Y}\left[\log f(y;\hat{\theta}_{k})\right]\right\} \tag{3.2.1.3}
\end{equation*}
can be asymptotically approximated by the dimension of the parameter space $k$. The AIC criterion is an
asymptotically unbiased estimation of the expected loglikelihood and can be expressed by the following expression
\begin{equation*}
AIC=-2\log f(y;\hat{\theta}_{k})+2k.
\end{equation*}

A series of the alternative information criteria based on the AIC information criterion have been proposed.
Sugiura (1978) proposed the AICc on the linear regression model.
Hurvich and Tsai (1989) extended the usage of the AICc on regression and time series models in small samples and
showed that the AIC and the AICc are asymptotically equivalent.
The AICc model selection performance will be more superior to AIC
under the circumstance of a finite sample size.
Schwarz(1978) proposed the Bayesian information criterion(BIC).
Hannan and Quinn (1979) proposed the HQ information criterion.

\subsubsection{The bootstrap extensions of the AIC criterion}
The motivation behind the bootstrap extensions of the AIC is to take advantage of the validated observation samples which will
be taken as a substitution toward the potential true population to fulfill the augmentation of the
training samples. The newly generated training samples can be used to realize the volatility of the parameter estimation,
which makes the calculation of the bias expression $(3.2.1.3)$ become possible.

Ishiguro (1997) proposed an information criterion known as EIC by using
the following bootstrap-based bias calculation expression,

\begin{equation*}
B_{1}=E_{Y^{b}}\left\{2\log f(y^{b}|\hat{\theta}^{b}_{k})-2\log f(y|\hat{\theta}^{b}_{k})\right\}. \tag{3.2.2.1}
\end{equation*}
Comparing the expression $(3.2.2.1)$ with the bias expression $(3.2.1.3)$, bias expression $(3.2.2.1)$ uses the probability density model
$f(y;\hat{\theta}_{k})$ as the generation mechanism of the potential population to generate bootstrapped
sample observation $y^{b}$. The volatility of the model parameter estimation can be fulfilled by generating
number of $B$ bootstrap-based maximum likelihood estimation $\hat{\theta}^{b}_{k}(i), i=1,...,B$ through
fitting the model $f(y|\theta_{k})$ to the number of $B$ bootstrapped samples $y^{b}(i), i=1,...,B$.
The bootstrap-based log-likelihood expression $\log f(y^{b}|\hat{\theta}^{b}_{k})$ is considered as
a substitution toward the log-likelihood $\log f(y|\hat{\theta}_{k})$.
The corresponding bootstrap-based expected log-likelihood is
calculated by the expression $\log f(y|\hat{\theta}_{k}^{b})$.

The difference between the bias expression $(3.2.1.3)$ and the bias expression $(3.2.2.1)$ is that the previous one
takes the assumed true probability model $f(y;\theta_{k_{0}})$
as the random generation mechanism of the potential unknown population; however, the later uses $y=(y_{1},...,y_{n})^{T}$
or its corresponding fitted probability model $f(y;\hat{\theta}_{k})$ to substitute the potential true population
so as to fulfill the augmentation of the bootstrap samples.
We refer such information criterion with the bias expression $(3.2.2.1)$ as the EIC1 criterion which can be expressed by the following expression
\begin{equation*}
EIC1=-2\log(y;\hat{\theta}_{k})+B_{1}.
\end{equation*}
According to the law of large numbers, the bias expression
$B_{1}$ can be approximated almost surely by the following expression
\begin{equation*}
\frac{1}{B}\sum\limits_{i=1}^{B}\left\{2\log f(y^{b}(i)|\hat{\theta}^{b}_{k}(i))-2\log f(y|\hat{\theta}^{b}_{k}(i))\right\},
\end{equation*}
where $B$ is the number of augmentation of the bootstrap sample $Y^{b}$.

Cavanaugh and Shumway (1997) proposed another similar bootstrap-based extension version of the AIC for the state-space model.
We refer it as the EIC2 criterion and its bias expression is
\begin{equation*}
B_{2}=2E_{Y^{b}}\left\{2\log f(y|\hat{\theta}_{k})-2\log f(y|\hat{\theta}_{k}^{b})\right\}. \tag{3.2.2.2}
\end{equation*}
We will refer to the information criterion with the bias expression $(3.2.2.2)$ as the EIC2 model selection criterion and it can be expressed by the following expression
\begin{equation*}
EIC2=-2\log(y;\hat{\theta}_{k})+B_{2}.
\end{equation*}

Shibata (1997) proposed another three bootstrap extensions of the AIC criterion and the corresponding bias expressions can be expressed as follows,
\begin{equation*}
B_{3}=2E_{Y^{b}}\left\{2\log f(y^{b}|\hat{\theta}^{b}_{k})-2\log f(y^{b}|\hat{\theta}_{k})\right\}
\end{equation*}

\begin{equation*}
B_{4}=2E_{Y^{b}}\left\{2\log f(y^{b}|\hat{\theta}_{k})-2\log f(y|\hat{\theta}_{k}^{b})\right\}
\end{equation*}

\begin{equation*}
B_{5}=2E_{Y^{b}}\left\{2\log f(y^{b}|\hat{\theta}^{b}_{k})-2\log f(y|\hat{\theta}_{k})\right\}
\end{equation*}
and we will refer to these three bootstrap-based model selection criteria as EIC3, EIC4 and EIC5 respectively which can be expressed as follows,

\begin{equation*}
EIC3=-2\log f(y;\hat{\theta}_{k})+B_{3}
\end{equation*}

\begin{equation*}
EIC4=-2\log f(y;\hat{\theta}_{k})+B_{4},
\end{equation*}

\begin{equation*}
EIC5=-2\log f(y;\hat{\theta}_{k})+B_{5}.
\end{equation*}

\subsubsection{The Bootstrap likelihood and cross-validation}

Pan, Wei (1999) introduced a model selection criterion
combined with the non-parametric bootstrap and the cross-validation:
bootstrap likelihood cross-validation (BCV). The BCV model selection criterion estimate the expected log-likelihood of the candidate model as following
\begin{equation*}
BCV=E_{Y^{b}}\left\{-2\log f(y^{-};\hat{\theta}_{k}^{b})\frac{n}{m^{*}}\right\}
\end{equation*}
where
$Y^{-}=Y-Y^{b}$, $Y^{b}$ is the non-parametric bootstrapped sample with the number of elements $n$, $m^{*}$ is the number of elements contained in $Y^{-}$,
$Y^{-}\bigcap Y^{b}=\emptyset$,
and $Y^{-}\bigcup Y^{b}=Y$.
Wei Pan (1999) introduced the following .632CV criterion following the $.632+$ rule proposed by Efron and Tibshirani (1997)

\begin{equation*}
632CV=0.368\left\{-2\log f(y|\hat{\theta}_{k})\right\}+0.632BCV.
\end{equation*}

\subsubsection{Bootstrap likelihood quasi-CV}
Bayer (2015) proposed a bootstrapped likelihood quasi-cross validation (BQCV)
model selection criterion which is similar to the bootstrapped likelihood cross-validation (BCV) criterion when
the generalized linear regression model is used to fit the dependent
variable with a beta distribution.

The BQCV criterion generates the training sample by using the empirical distribution $\hat{F}$ estimated at $\hat{\theta}_{k}$. The newly generated training sample $Y^{b}$ combining with the validated sample $Y$ constitutes the final observation sample.
Comparing with the real cross validation, the BQCV criterion uses the bootstrapped samples $y^{b}$ as the training sample and take the validated sample
$Y$ as the testing sample to estimate the expected log-likelihood.
The BQCV criterion can be expressed as

\begin{equation*}
BQCV=E_{Y^{b}}\left\{-2\log f(y|\hat{\theta}^{b}_{k})\right\}.
\end{equation*}
Bayer (2015) also proposed the corresponding 632BQCV model selection criterion which can be expressed by the following expression

\begin{equation*}
632QCV=0.368\left\{-2\log f(y;\hat{\theta}_{k})\right\}+0.632BQCV.
\end{equation*}

\section{Tobit regression model and its sample augmentation mechanism}

\subsection{Tobit regression model and its maximum likelihood}
The Tobit regression model, which is a generalization of the Probit regression model, was proposed by  Tobin,J.(1958)\cite{tobin1958estimation} when he analyzed the data
of the household expenditure on the durable goods. Takeshi Amemiya(1973) \cite{amemiya1973regression}
proved the consistency and the asymptotic normality of the maximum likelihood estimation
of the Tobit regression model. Tobit regression model can be described as follows:

\begin{equation*}
y_{i}=
\begin{cases}
y_{i}^{*}  & \text{if $y_{i}^{*}=x_{i}^{T}\beta+\varepsilon_{i}>0$}\\
0      & \text{if $y_{i}^{*}=x_{i}^{T}\beta+\varepsilon_{i}\leq0$}, \tag{4.1.1}
\end{cases}
\end{equation*}
where $i, i=1,...,n$ denotes the $i$th observation individual, $y_{i}$ is the realization of the interesting response variable and
$x_{i}$ is the corresponding covariate,
$\beta$ is the unknown parameter vector or the contribution effects on $x_{i}$. The disturbance terms
$\varepsilon_{i}, i=1,...,n$ are independently identically distributed (i.i.d.) Gaussian white noise with zero mean
and unknown variance $\sigma^{2}$. The regression model will only contain the intercept term and the disturbance term when $x_{i}=1, i=1,...,n$.

The probability density of the positive observation $y_{i},i\in\left\{i:y_{i}>0\right\}$ under the model $(4.1.1)$ can be expressed as
\begin{equation*}
f(y_{i})=\phi_{0,\sigma^{2}}(y_{i}-x_{i}^{T}\beta)=\frac{1}{\sqrt{2\pi}\sigma}\exp\left\{-\frac{(y_{i}-x_{i}^{T}\beta)^{2}}{2\sigma^{2}}\right\} ,
\end{equation*}
where $\phi_{0,\sigma^{2}}$ denotes the probability density function of the normal distribution with mean $0$ and variance $\sigma^{2}$.
The probability of zero value observation $y_{i}=0, i\in\left\{i:y_{i}=0\right\}$ under the model $(4.1.1)$ can be expressed as

\begin{align*}
P(x_{i}^{T}\beta+\varepsilon_{i}\leq0)&=1-\int_{-\infty}^{x_{i}^{T}\beta}\frac{1}{\sqrt{2\pi}\sigma}\exp\left(-\frac{\varepsilon_{i}^{2}}{2\sigma^{2}}\right)d\varepsilon_{i}\\
&\overset{\frac{\varepsilon_{i}}{\sigma}=\eta_{i}}{=}1-\int_{-\infty}^{x_{i}^{T}\beta}\frac{1}{\sqrt{2\pi}}\exp\left(-\frac{\eta_{i}^{2}}{2}\right)d\eta_{i}
&=1-\Phi\left(\frac{x_{i}^{T}\beta}{\sigma}\right),
\end{align*}
where $\Phi$ denotes the standard normal cumulative distribution function.

The observation likelihood function of the model $(4.1.1)$ can be expressed by the following expression accordingly
\begin{align*}
\left\{\prod_{\left\{i:y_{i}>0\right\}}\left(\frac{1}{\sqrt{2\pi}\sigma}\exp\left[-\frac{(y_{i}-x_{i}^{T}\beta)^{2}}{2\sigma^{2}}\right]\right)\right\}\cdot
\left\{\prod_{\left\{i:y_{i}=0\right\}}\left[1-\Phi\left(\frac{x_{i}^{T}\beta}{\sigma}\right)\right]\right\},
\end{align*}
and loglikelihood of the Tobit regression model $(4.1.1)$ is
\begin{equation*}
u\log\frac{1}{\sqrt{2\pi}\sigma}-\frac{1}{2{\sigma^{2}}}\sum_{\left\{i:y_{i}>0\right\}}(y_{i}-x_{i}^{T}\beta)^{2}+\sum_{\left\{i:y_{i}=0\right\}}\log \left[1-\Phi\left(\frac{x_{i}^{T}\beta}{\sigma}\right)\right],
\end{equation*}
where $u$ is the number of positive observation individuals.

\subsection{The bootstrapped sampling mechanism of the Tobit regression model}
In this section, we mainly consider three kinds of bootstrap sampling mechanisms of the Tobit regression model which are parametric bootstrap, non-parametric bootstrap and the combination of the parametric and nonparametric bootstrap sampling mechanism. The final newly generated bootstrap samples will be considered as the
training samples coming from the potential unknown true population.
\subsubsection{Nonparametric Bootstrap}
The nonparametric bootstrap sampling technique is a data augmentation methodology by generating the bootstrap samples from the empirical distribution
function which is acquired by assigning the weight $1/n$ to each observation individual $y_{i}, i=1,...,n$.
The nonparametric bootstrap sampling methodology does not have to require making any model assumption about the potential unknown population.
Therefore, the nonparametric sample augmentation mechanism is an extremely straightforward and effective
sample generation mechanism.

\subsubsection{Parametric Bootstrap}
The parametric bootstrap sampling mechanism is also a commonly used data augmentation technique because
the estimation of the model parameter contains the information of the potential unknown population.
The parametric bootstrap generation mechanism of the Tobit regression model (4.1.1) can be described as follows,

\begin{equation*}
y_{i}^{b}=\left\{
        \begin{array}{ll}
          y_{i}^{b^{*}}, & \hbox{if $y_{i}^{b^{*}}=x_{i}^{T}\hat{\beta}+\hat{\varepsilon}_{i}>0$;} \\
          0, & \hbox{if $y_{i}^{b^{*}}=x_{i}^{T}\hat{\beta}+\hat{\varepsilon}_{i}\leq0$},
        \end{array}
      \right.
\end{equation*}
where $y_{i}^{*}$ is the $i$th bootstrap observation individual, $\hat{\beta}$ is the maximum likelihood estimation based on the
number of $n$ validated observations, $\hat{\sigma}$ is the maximum likelihood estimation of the model disturbance term $\sigma$ and $\hat{\varepsilon}_{i}\sim N(0,\hat{\sigma}^{2})$.

\subsubsection{Integration of the nonparametric and the parametric bootstrap sampling mechanisms}
The nonparametric bootstrap sampling methodology does not depend on any specific model assumption and the bootstrap samples completely
come from the empirical distribution.
Meanwhile, the parametric bootstrap sampling mechanism can sufficiently take advantage of the model parameter information to generate bootstrap samples.
Therefore, it is natural to integrate both the nonparametric bootstrap and the parametric bootstrap sampling mechanisms to increase the
variability of the random observation samples.
The random bootstrap observation of the combination methodology can be expressed as follows:

\begin{equation*}
y_{i}^{b}=\left\{
        \begin{array}{ll}
          y_{i}^{b^{*}}, & \hbox{if $y_{i}^{b^{*}}=x_{i}^{b^{T}}\hat{\beta}+\hat{\varepsilon}_{i}>0$;} \\
          0, & \hbox{if $y_{i}^{b^{*}}=x_{i}^{b^{T}}\hat{\beta}+\hat{\varepsilon}_{i}\leq0$,}
        \end{array}
      \right.
\end{equation*}
where $x_{i}^{b}, i=1,...,n$ is the covariate generated by the nonparametric bootstrap sampling mechanism, $\hat{\varepsilon}_{i}\sim N(0,\hat{\sigma}^{2})$,
$\hat{\beta}$ and $\hat{\sigma}^2$ are the maximum likelihood estimations of the model parameter $\beta$ and the variance of the
model disturbance $\sigma^{2}$ respectively based on the number of $n$ validated observation samples.

\section{Simulation Study}
In this simulation study section, we use the Monte Carlo simulation experimentation to demonstrate the performances of the model selection
of different kinds of model selection criteria on the Tobit regression model. To save the cost of the intensive computation and
sufficiently demonstrate the model selection performances of different kinds of model selection criteria, all the explanatory variables will be assigned
by the same marginal distribution and the correlation coefficient between any two different explanatory variables will be set as the same constant.

The potential possible explanatory vector is $X=(1,X_{1},...,X_{p})^{T}$, where the random vector $(X_{1},...,X_{p})^{T}$ follows the $p$-dimension Gaussian distribution
with the zero mean and the variance-covariance matrix $\Sigma_{p\times p}$. The corresponding regression coefficient
is $\beta=(\beta_{0},\beta_{1},...,\beta_{p})^{T}$, where $\beta_{0}$ is the intercept term which is used to adjust the response variable
censoring rate.

In this simulation study section, we set the maximum allowable number of the explanatory variables $p=8$.
Then the maximum allowable dimension of the candidate model class of the Tobit regression model $\mathcal{F}=\left\{\mathcal{F}(k):1\leq k\leq m\right\}$
is $m=p+2=10$ which is the summation of the $(p+1)$-dimension
regression coefficient $\beta$ and one dimension variance of the disturbance term $\sigma^{2}$.
Meanwhile, we set the variance of the disturbance term $\sigma^{2}=1$ and
the true regression coefficient between the response variable $Y_{i}$ and $X_{i}$ is
$\beta=(\beta_{0},0.1,0.2,0.3,0.4,0,0,0,0)^{T}$, where the random vector $X_{i}=(1,X_{i1},...,X_{ip})^{T}$ is the $i$th random observation of $X=(1,X_{1},...,X_{p})^{T}$
and $i=1,...,n$.
The variance-covariance matrix of $(X_{1},...,X_{p})^{T}$ is assigned by setting $\Sigma_{ij}=1$ for $i=j$, $\Sigma_{ij}=0.3$ for $i\neq j$, where $\Sigma_{ij}$ denotes the component of the $i$th row, the $j$th column of the variance and covariance matrix $\Sigma_{p\times p}$, $i,j=1,...,p$.
The number of the bootstrapped sample augmentation will be set as $B=200$ in this simulation study section.

The observation response $y_{i}$ is generated as follows,
\begin{equation*}
y_{i}=
\begin{cases}
y_{i}^{*}   & \text{if $y_{i}^{*}=\beta_{0}+0.1x_{i1}+0.2x_{i2}+0.3x_{i3}+0.4x_{i4}+\varepsilon_{i}>0$}\\
0           & \text{if $y_{i}^{*}=\beta_{0}+0.1x_{i1}+0.2x_{i2}+0.3x_{i3}+0.4x_{i4}+\varepsilon_{i}\leq0$},
\end{cases}
\end{equation*}
where $i=1,...,n$, $\varepsilon_{i}\sim N(0,1)$, $y_{i}^{*}$ is the realization of $Y_{i}^{*}$ which follows the Gaussian distribution with
the expectation $E(Y_{i}^{*})=E(X^{T}\beta)=\beta_{0}$ and the variance
Var$(Y_{i}^{*})=$Var$(X^{T}\beta+\varepsilon_{i})=\beta_{[2:5]\times1}^{T}\Sigma_{[1:4]\times[1:4]}\beta_{[2:5]\times1}+1$, where $\beta_{[2:5]\times1}=(0.1,0.2,0.3,0.4)^{T}$
and $\Sigma_{[1:4]\times[1:4]}$ denotes the matrix with the first four rows and the first four columns of $\Sigma_{p\times p}$.
The censoring rate of $Y_{i}$ is $P(N(\beta_{0},\beta^{T}_{[2:5]\times1}\Sigma_{[1:4]\times[1:4]}\beta_{[2:5]\times1}+1)<0)$, where the notation expression $N(\beta_{0},\beta^{T}_{[2:5]\times1}\Sigma_{[1:4]\times[1:4]}\beta_{[2:5]\times1}+1)$ denotes
the Gaussian distribution random variable with mean $\beta_{0}$ and the variance $\beta_{[2:5]\times1}^{T}\Sigma_{[1:4]\times[1:4]}\beta_{[2:5]\times1}+1$.

There are a total number of ten candidate model families which can be expressed by the following expression
\begin{equation*}
\mathcal{F}(k)=\left\{f(y;\beta_{0},\beta_{1},...,\beta_{k},\sigma^{2}): (\beta_{0},\beta_{1},...,\beta_{d},\sigma^{2})\in  R^{d+1}\times R^{+} \right\},
\end{equation*}
More specifically, the total number of ten candidate model families can be demonstrated as follows:

$\mathcal{F}(1)=\left\{f(y;\sigma^{2}): \sigma^{2}\in R^{+} \right\}$

$\mathcal{F}(2)=\left\{f(y;\beta_{0},\sigma^{2}): (\beta_{0},\sigma^{2})\in R\times R^{+} \right\}$

$\mathcal{F}(3)=\left\{f(y;\beta_{0},\beta_{1},\sigma^{2}): (\beta_{0},\beta_{1},\sigma^{2})\in R^{2}\times R^{+}\right\}$

$\mathcal{F}(4)=\left\{f(y;\beta_{0},\beta_{1},\beta_{2},\sigma^{2}): (\beta_{0},\beta_{1},\beta_{2},\sigma^{2})\in  R^{3}\times R^{+}\right\}$

$\mathcal{F}(5)=\left\{f(y;\beta_{0},\beta_{1},\beta_{2},\beta_{3},\sigma^{2}): (\beta_{0}, \beta_{1},\beta_{2},\beta_{3},\sigma^{2})\in  R^{4}\times R^{+}\right\}$

$\mathcal{F^{*}}(6)=\left\{f(y;\beta_{0}, \beta_{1}, \beta_{2},\beta_{3},\beta_{4},\sigma^{2}): (\beta_{0},\beta_{1},\beta_{2},\beta_{3},\beta_{4},\sigma^{2})\in R^{5}\times R^{+}\right\}$

\quad\vdots

$\mathcal{F}(10)=\left\{f(y;\beta_{0},\beta_{1},\beta_{2},\beta_{3}...,\beta_{8},\sigma^{2}): (\beta_{0}, \beta_{1},...,\beta_{8},\sigma^{2})\in R^{9}\times R^{+}\right\}.$

From the perspective of the variable selection of the regression model, the most simple regression model family is
\begin{equation*}
\mathcal{F}(2)=\left\{f(y;\beta_{0},\sigma^{2}):(\beta_{0},\sigma^{2})\in R\times R^{+}\right\}.
\end{equation*}
If we use the notation $d, 0\leq d\leq 8$ to denote the number of explanatory variables got involved into the candidate model,
the number of explanatory variables got involved in the candidate model family $\mathcal{F}(k)$ is $d=k-2$ for $k\geq2$.
Specifically, the candidate model family $\mathcal{F}(2)$ is corresponding to the most simple regression model which only includes
the intercept term $\beta_{0}$ and the error term $\varepsilon_{i}$ with the disturbance variance $\sigma^{2}$.
Accordingly, the number of explanatory variables got involved in the candidate model class $\mathcal{F}(2)$ is $0$.
The true model belongs to the candidate model family $\mathcal{F}(6)$ with the
true number of explanatory variables $d_{0}=6-2=4$. The objective of the model selection is to select the correct number of variables $d_{0}=4$ from the candidate model class
$\left\{\mathcal{F}(2),...,\mathcal{F}(10)\right\}$.

The performances of the model selection are demonstrated from the Table 1 to the Table 4, where the subscripts of the names of the criteria pb, np, and npp stand for
parametric bootstrap, nonparametric bootstrap and the combination of the nonparametric and parametric bootstrap respectively.

\setlength{\tabcolsep}{0.8mm}{
\begin{table*}[h]
\scriptsize
\centering
\caption{The identification frequency on the Tobit regression model with the interesting variable censoring rate $0.75$,
the variance of the Gaussian white noise $\sigma^{2}=1$, the true number of explanatory variables $d_{0}=4$, the regression coefficient
$\beta=(-0.84,0.1,0.2,0.3,0.4,0,0,0,0)^{T}$ and the number of Monte Carlo experimentation is $500$}
\label{Tab04}
\begin{tabular}{lcccccccccccc}
  \toprule
\multirow{2}{*}{}&\multicolumn{3}{c}{$n=100$}&\multicolumn{3}{c}{$n=120$}&\multicolumn{3}{c}{$n=150$}&\multicolumn{3}{c}{$n=200$}\\
\cmidrule(r){2-4}\cmidrule(r){5-7}\cmidrule(r){8-10}\cmidrule(r){11-13}
&$<d_{0}$&$=d_{0}$&$>k_{0}$&$<d_{0}$&$=d_{0}$&$>d_{0}$&$<d_{0}$&$=d_{0}$&$>d_{0}$&$<d_{0}$&$=d_{0}$&$>d_{0}$\\
\midrule
AIC               &53&308&139          &32&343&125          &23&342&135              &2&358&140                                  \\
BIC               &208&279&13          &161&320&19          &113&364&23              &56&431&13                           \\
AICc              &62&318&120          &42&351&107          &29&348&123              &2&372&126                                  \\
HQ                &37&263&200          &22&294&184          &12&306&182              &1&321&178                                  \\
$\rm EIC1_{np}$   &86&333&81           &55&369&76          &33&358&109              &4&373&123                                \\
$\rm EIC1_{pb}$   &55&153&292          &26&131&343          &16&118&366              &6&107&387                                \\
$\rm EIC1_{npp}$  &115&237&148         &77&213&210          &37&208&255              &16&185&299                                  \\
$\rm EIC2_{np}$   &126&330&44          &73&367&60          &40&367&93               &9&391&100                                  \\
$\rm EIC2_{pb}$   &37&205&258          &19&176&305          &14&159&327              &4&141&355                               \\
$\rm EIC2_{npp}$  &122&340&38          &72&368&60          &39&392&69               &9&400&91                                \\
$\rm EIC3_{np}$   &41&177&282          &23&192&285          &15&205&280              &2&186&312                                \\
$\rm EIC3_{pb}$   &45&219&236          &28&256&216          &23&267&210              &2&294&204                              \\
$\rm EIC3_{npp}$  &46&222&232          &35&244&221          &18&259&223              &2&283&215                                \\
$\rm EIC4_{np}$   &101&302&97          &64&328&108          &31&318&151              &7&310&183                                \\
$\rm EIC4_{pb}$   &373&23&104          &369&15&116          &384&13&103              &401&12&87                                \\
$\rm EIC4_{npp}$  &463&16&21           &458&16&20          &446&21&33               &460&10&30                                    \\
$\rm EIC5_{np}$   &46&236&218          &33&281&186          &21&287&192              &3&295&202                                  \\
$\rm EIC5_{pb}$   &404&22&74           &420&14&66          &420&10&70               &439&13&48                                   \\
$\rm EIC5_{npp}$  &406&21&73           &416&12&72          &418&9&73                &442&10&48                                   \\
$\rm BCV$        &250&241&9           &160&322&18          &87&\bf{382}&31          &30&\bf{432}&38                                    \\
$\rm CV632$       &115&\bf{344}&41     &77&\bf{373}&50          &38&380&82               &10&395&95                                         \\
$\rm BQCV$        &10&147&343          &7&121&372          &1&96&403                &1&89&410                                              \\
$\rm QCV632$      &5&100&395           &622&85&413          &0&60&440                &0&51&449                                          \\
\bottomrule
\end{tabular}
\end{table*}}

\begin{table*}[h]
\scriptsize
\centering
\caption{The identification frequency on the Tobit regression model with the interesting variable censoring rate $0.7$,
the variance of the Gaussian white noise $\sigma^{2}=1$, the true number of explanatory variables $d_{0}=4$, the regression coefficient
$\beta=(-0.65,0.1,0.2,0.3,0.4,0,0,0,0)^{T}$ and the number of Monte Carlo experiments is set as $500$}
\label{Tab04}
\begin{tabular}{lcccccccccccc}
  \toprule
\multirow{2}{*}{}&\multicolumn{3}{c}{$n=100$}&\multicolumn{3}{c}{$n=120$}&\multicolumn{3}{c}{$n=150$}&\multicolumn{3}{c}{$n=200$}\\
\cmidrule(r){2-4}\cmidrule(r){5-7}\cmidrule(r){8-10}\cmidrule(r){11-13}
&$<d_{0}$&$=d_{0}$&$>d_{0}$&$<d_{0}$&$=d_{0}$&$>d_{0}$&$<d_{0}$&$=d_{0}$&$>d_{0}$&$<d_{0}$&$=d_{0}$&$>d_{0}$\\
\midrule
AIC               &57&315&128          &16&357&127          &10&362&128              &3&371&126                                  \\
BIC               &184&295&21          &135&345&20          &86&\bf{395}&19              &42&\bf{450}&8                           \\
AICc              &64&321&115          &24&361&115          &14&371&115              &3&382&115                                  \\
HQ                &38&268&194          &11&312&177          &6&328&166              &2&330&168                                  \\
$\rm EIC1_{np}$   &89&333&78           &28&383&89          &20&379&101              &3&393&104                                \\
$\rm EIC1_{pb}$   &67&134&299          &41&158&301          &26&148&326              &7&122&371                                \\
$\rm EIC1_{npp}$  &151&199&150         &84&245&171          &56&218&226              &20&218&262                                  \\
$\rm EIC2_{np}$   &106&333&61          &40&391&69          &21&392&87              &4&404&92                                  \\
$\rm EIC2_{pb}$   &36&159&305          &15&177&308          &19&152&339              &3&128&369                               \\
$\rm EIC2_{npp}$  &109&\bf{337}&54     &44&392&64          &27&389&84              &7&408&85                                \\
$\rm EIC3_{np}$   &41&206&253          &9&232&259          &6&250&244              &2&231&267                                \\
$\rm EIC3_{pb}$   &52&270&178          &19&306&175          &11&325&164              &2&326&172                              \\
$\rm EIC3_{npp}$  &52&269&179          &17&308&175          &12&315&173              &2&315&183                                \\
$\rm EIC4_{np}$   &106&303&91          &40&356&104          &23&328&149              &3&328&169                                \\
$\rm EIC4_{pb}$   &375&18&107          &376&17&107          &395&14&91              &417&9&74                                \\
$\rm EIC4_{npp}$  &455&21&24           &459&21&20          &458&16&26              &473&6&21                                    \\
$\rm EIC5_{np}$   &56&281&163          &19&327&154          &13&331&156              &2&346&152                                  \\
$\rm EIC5_{pb}$   &430&20&50           &423&21&56          &447&16&37              &461&4&35                                   \\
$\rm EIC5_{npp}$  &430&19&51           &423&21&56          &444&18&38              &460&3&37                                   \\
$\rm BCV$        &206&276&18          &109&365&26          &63&392&45              &17&450&33                                    \\
$\rm CV632$       &111&335&54          &38&\bf{397}&65          &22&394&84              &3&410&87                                         \\
$\rm BQCV$        &8&113&379           &4&106&390          &0&93&407              &0&72&428                                              \\
$\rm QCV632$      &2&72&426            &0&62&438          &0&65&435              &0&39&461                                          \\
\bottomrule
\end{tabular}
\end{table*}

\begin{table*}[h]
\scriptsize
\centering
\caption{The identification frequency on the Tobit regression model with the interesting variable censoring rate $0.6$,
the variance of the Gaussian white noise $\sigma^{2}=1$, the true number of explanatory variables $d_{0}=4$, the regression coefficient
$\beta=(-0.28,0.1,0.2,0.3,0.4,0,0,0,0)^{T}$ and the number of Monte Carlo experiments is set as $500$}
\label{Tab04}
\begin{tabular}{lcccccccccccc}
  \toprule
\multirow{2}{*}{}&\multicolumn{3}{c}{$n=100$}&\multicolumn{3}{c}{$n=120$}&\multicolumn{3}{c}{$n=150$}&\multicolumn{3}{c}{$n=200$}\\
\cmidrule(r){2-4}\cmidrule(r){5-7}\cmidrule(r){8-10}\cmidrule(r){11-13}
&$<d_{0}$&$=d_{0}$&$>d_{0}$&$<d_{0}$&$=d_{0}$&$>d_{0}$&$<d_{0}$&$=d_{0}$&$>d_{0}$&$<d_{0}$&$=d_{0}$&$>d_{0}$\\
\midrule
AIC               &38&319&143           &11&343&146          &6&359&135              &2&362&136                                  \\
BIC               &139&343&18           &74&\bf{404}&22          &54&\bf{433}&13              &16&\bf{466}&18                           \\
AICc              &39&334&127           &15&352&133          &6&379&115              &3&376&121                                  \\
HQ                &27&263&210           &9&290&201          &3&314&183              &2&312&186                                  \\
$\rm EIC1_{np}$   &51&366&83           &17&371&112          &9&381&110              &3&381&116                                \\
$\rm EIC1_{pb}$   &49&152&299           &22&164&314          &14&155&331              &3&142&355                                \\
$\rm EIC1_{npp}$  &106&251&143           &61&257&182          &34&259&207              &11&249&240                                  \\
$\rm EIC2_{np}$   &61&378&61           &26&383&91          &14&400&86              &3&389&108                                  \\
$\rm EIC2_{pb}$   &28&145&327           &9&130&361          &13&108&379              &2&109&389                               \\
$\rm EIC2_{npp}$  &58&378&64           &22&392&86          &14&386&100              &2&392&106                               \\
$\rm EIC3_{np}$   &37&257&206           &9&272&219          &7&271&222              &3&264&233                                \\
$\rm EIC3_{pb}$   &38&305&157           &14&319&167          &5&342&153              &3&349&148                              \\
$\rm EIC3_{npp}$  &39&299&162           &13&305&182          &10&332&158              &2&347&151                                \\
$\rm EIC4_{np}$   &60&362&78           &28&372&100          &14&373&113              &2&346&152                                \\
$\rm EIC4_{pb}$   &162&56&282           &180&53&267          &148&51&301              &134&53&313                                \\
$\rm EIC4_{npp}$  &305&111&84           &287&102&111          &273&100&127              &246&97&157                                    \\
$\rm EIC5_{np}$   &44&313&143           &16&330&154          &7&348&145              &2&356&142                                  \\
$\rm EIC5_{pb}$   &260&85&155           &253&76&171          &249&78&173              &224&84&192                                   \\
$\rm EIC5_{npp}$  &257&86&157           &255&82&163          &246&86&168              &223&79&198                                   \\
$\rm BCV$        &128&346&26           &63&403&34          &37&430&33              &5&451&44                                    \\
$\rm CV632$       &62&\bf{381}&57      &25&379&96          &14&398&88              &3&388&109                                         \\
$\rm BQCV$        &1&89&410           &1&70&429          &0&58&442              &0&61&439                                              \\
$\rm QCV632$      &0&46&454           &1&41&458          &0&27&473              &0&31&469                                          \\
\bottomrule
\end{tabular}
\end{table*}

\begin{table*}[h]
\scriptsize
\centering
\caption{The identification frequency on the Tobit regression model with the interesting variable censoring rate $0.5$,
the variance of the Gaussian white noise $\sigma^{2}=1$, the true number of explanatory variables $d_{0}=4$, the regression coefficient
$\beta=(0,0.1,0.2,0.3,0.4,0,0,0,0)^{T}$ and the number of Monte Carlo experiments is set as $500$}
\label{Tab04}
\begin{tabular}{lcccccccccccc}
  \toprule
\multirow{2}{*}{}&\multicolumn{3}{c}{$n=100$}&\multicolumn{3}{c}{$n=120$}&\multicolumn{3}{c}{$n=150$}&\multicolumn{3}{c}{$n=200$}\\
\cmidrule(r){2-4}\cmidrule(r){5-7}\cmidrule(r){8-10}\cmidrule(r){11-13}
&$<d_{0}$&$=d_{0}$&$>d_{0}$&$<d_{0}$&$=d_{0}$&$>d_{0}$&$<d_{0}$&$=d_{0}$&$>d_{0}$&$<d_{0}$&$=d_{0}$&$>d_{0}$\\
\midrule
AIC               &10&344&146           &13&362&125          &1&355&144              &0&365&135                                  \\
BIC               &100&375&25           &69&412&19          &33&\bf{449}&18              &11&\bf{478}&11                           \\
AICc              &16&361&123           &19&364&117          &1&366&133              &1&376&123                                  \\
HQ                &6&287&207           &8&301&191          &1&301&198              &0&327&173                                  \\
$\rm EIC1_{np}$   &26&374&100           &22&388&90          &2&382&116              &2&383&115                                \\
$\rm EIC1_{pb}$   &9&191&300           &15&177&308          &2&155&343              &0&138&362                                \\
$\rm EIC1_{npp}$  &33&297&170           &22&309&169          &8&288&204              &1&252&247                                  \\
$\rm EIC2_{np}$   &34&397&69           &27&400&73          &3&396&101              &2&400&98                                  \\
$\rm EIC2_{pb}$   &18&130&352           &15&110&375          &7&94&399              &0&87&413                               \\
$\rm EIC2_{npp}$  &31&377&92           &27&385&88          &2&382&116              &1&387&112                               \\
$\rm EIC3_{np}$   &11&308&181           &16&280&204          &1&272&227              &1&269&230                                \\
$\rm EIC3_{pb}$   &14&339&147           &19&333&148          &2&341&157              &0&341&159                              \\
$\rm EIC3_{npp}$  &18&336&146           &16&339&145          &1&344&155              &0&341&159                                \\
$\rm EIC4_{np}$   &33&390&77           &25&389&86          &2&370&128              &1&362&137                                \\
$\rm EIC4_{pb}$   &26&92&382           &19&90&391          &12&87&401              &1&70&429                                \\
$\rm EIC4_{npp}$  &80&239&181           &60&247&193          &38&215&247              &8&182&310                                    \\
$\rm EIC5_{np}$   &19&340&141           &15&350&135          &1&359&140              &2&345&153                                  \\
$\rm EIC5_{pb}$   &47&196&257           &46&211&243          &24&185&291              &6&163&331                                   \\
$\rm EIC5_{npp}$  &51&199&250           &40&215&245          &25&186&289              &7&154&339                                   \\
$\rm BCV$        &81&385&34           &60&\bf{414}&26          &15&436&49              &5&451&44                                    \\
$\rm CV632$       &34&\bf{392}&74           &24&403&73          &3&393&104              &2&393&105                                         \\
$\rm BQCV$        &1&71&428           &2&58&440          &1&58&441              &0&48&452                                              \\
$\rm QCV632$      &1&30&469           &0&29&471          &0&27&473              &0&23&477                                          \\
\bottomrule
\end{tabular}
\end{table*}

The simulation results demonstrate that the non-parametric bootstrap performance will be superior to the other bootstrap sampling mechanism for the EIC1, EIC4,
and EIC5 model selection criteria. As for the EIC2 model selection criterion, the performance of $EIC2_{npp}$ will be superior to the $EIC2_{p}$ and the
$EIC2_{np}$. However, the model selection performance of the $EIC3_{np}$ will be inferior to the $EIC3_{pb}$ and $EIC3_{npp}$.
When the observation information is adequate, the BIC and BCV criteria are becoming competitive than the other model selection criteria.
However, the CV632 criterion is becoming superior to the others when the observation information is becoming inadequate.

Meanwhile, to clearly demonstrate the performance of the model selection, a risk function is defined by the expression
$E\left\{I(\hat{d}={d}_{0})\right\}=P(\hat{d}=d_{0})$
which can be approximately calculated by the expression
$\left\{\sum\limits_{i=1}^{M}I(\hat{d}_{i}=d_{0})\right\}/M$,
where $\hat{d}$ denotes the estimation of the number of explanatory variables of the $i$th Monte Carlo experiment, $M$ is the number of the Monte Carlo
experimentation and $I(\cdot)$ stands for the indicator function. The graph of the risk functions of the BIC, BCV, and CV632 criteria are demonstrated from the graphs (a)-(h).

\begin{figure}[htbp]
\subfigure[]{
\centering
\includegraphics[width=5.8cm]{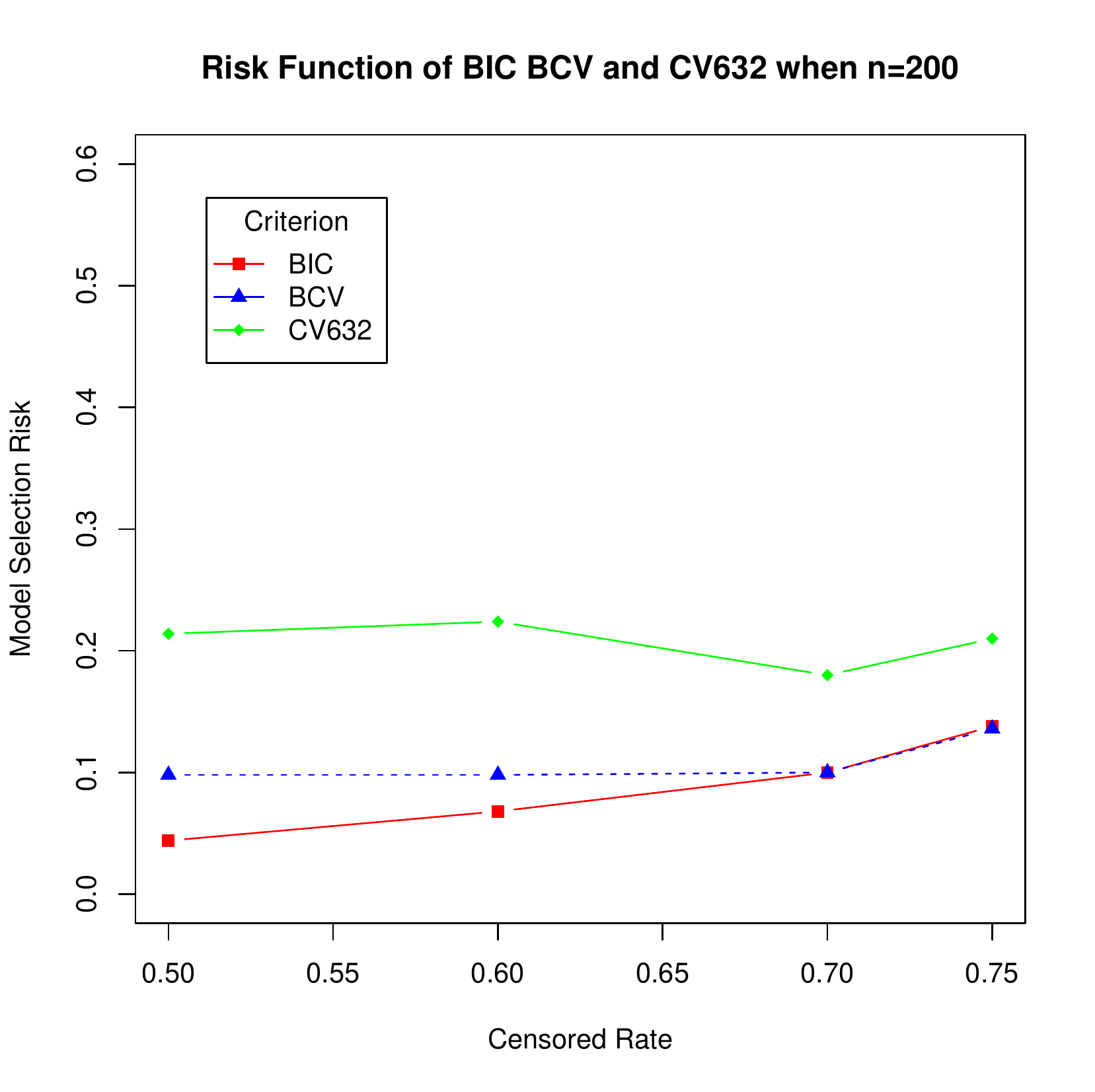}}
\subfigure[]{
\centering
\includegraphics[width=5.8cm]{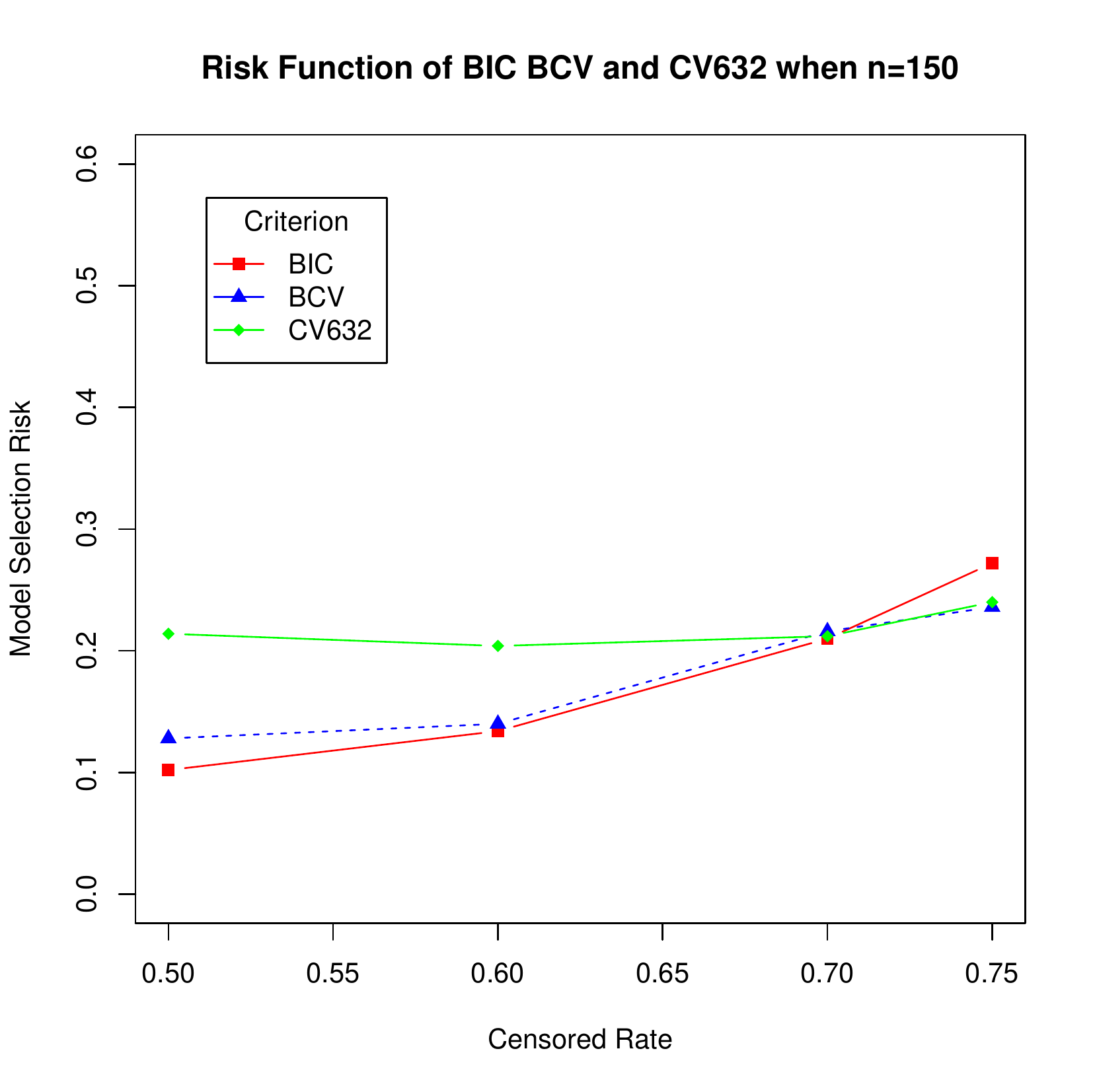}}
\subfigure[]{
\centering
\includegraphics[width=5.8cm]{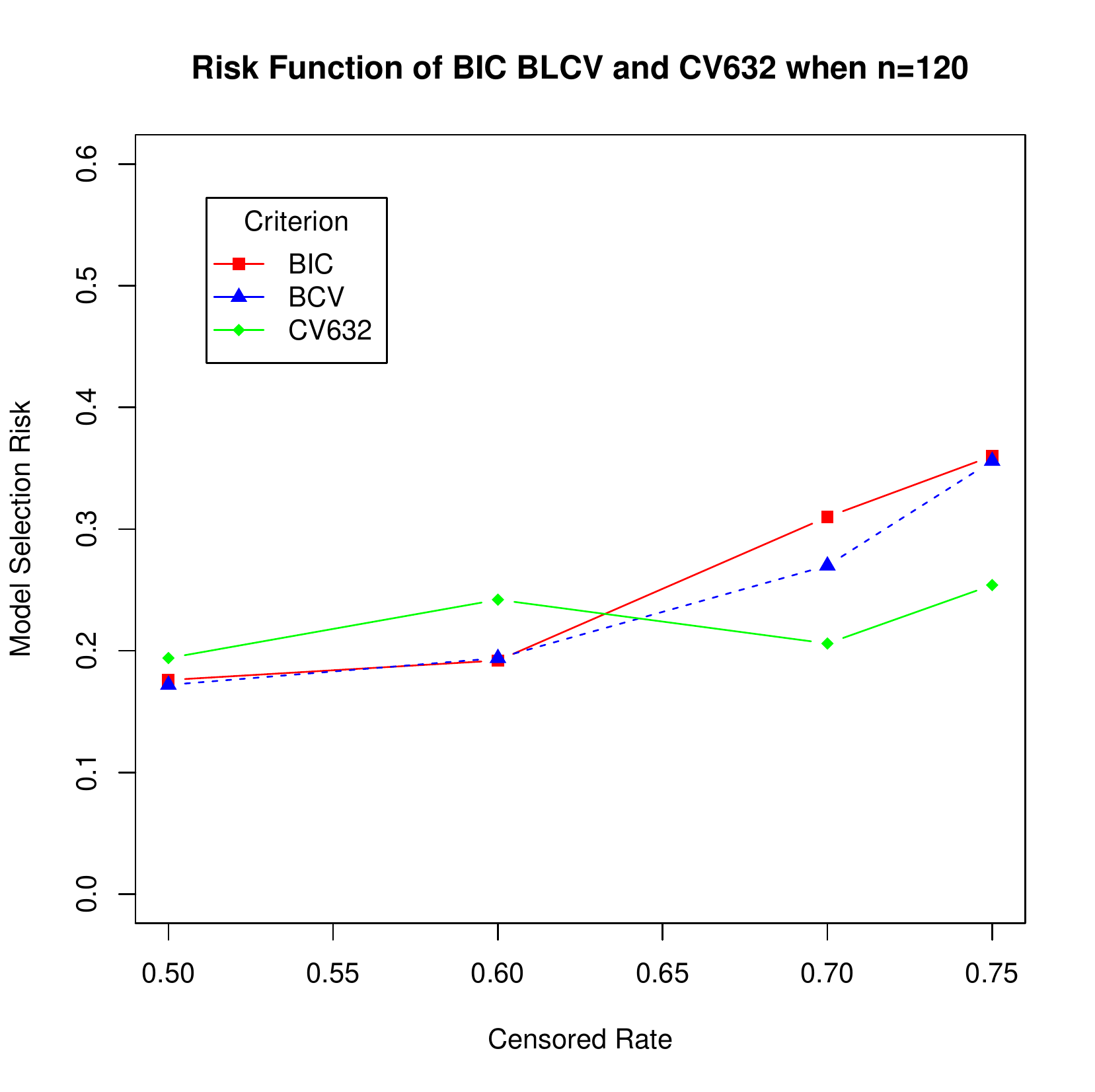}}
\subfigure[]{
\centering
\includegraphics[width=5.8cm]{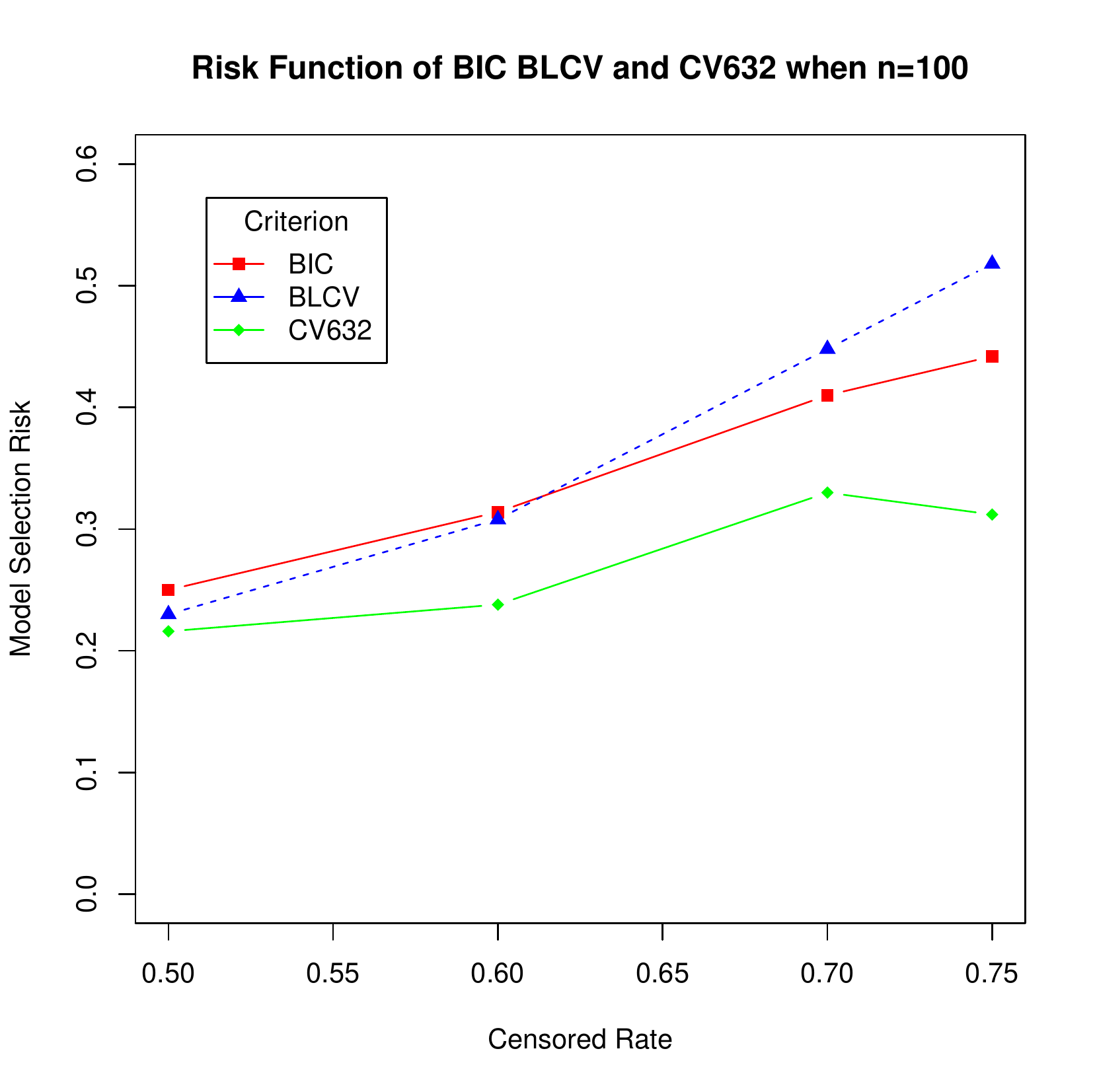}}
\end{figure}

\begin{figure}[htbp]
\subfigure[]{
\centering
\includegraphics[width=5.8cm]{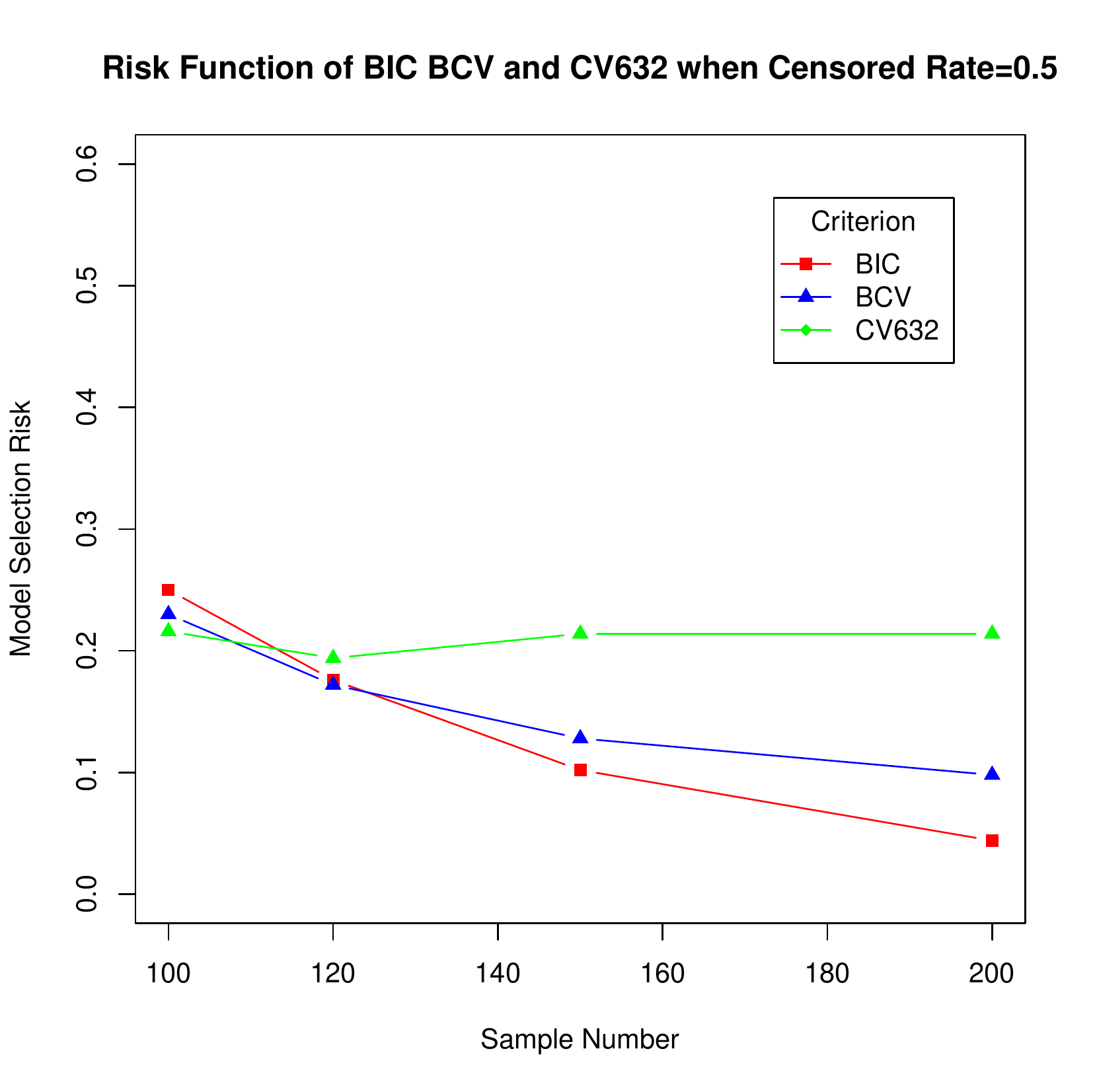}}
\subfigure[]{
\centering
\includegraphics[width=5.8cm]{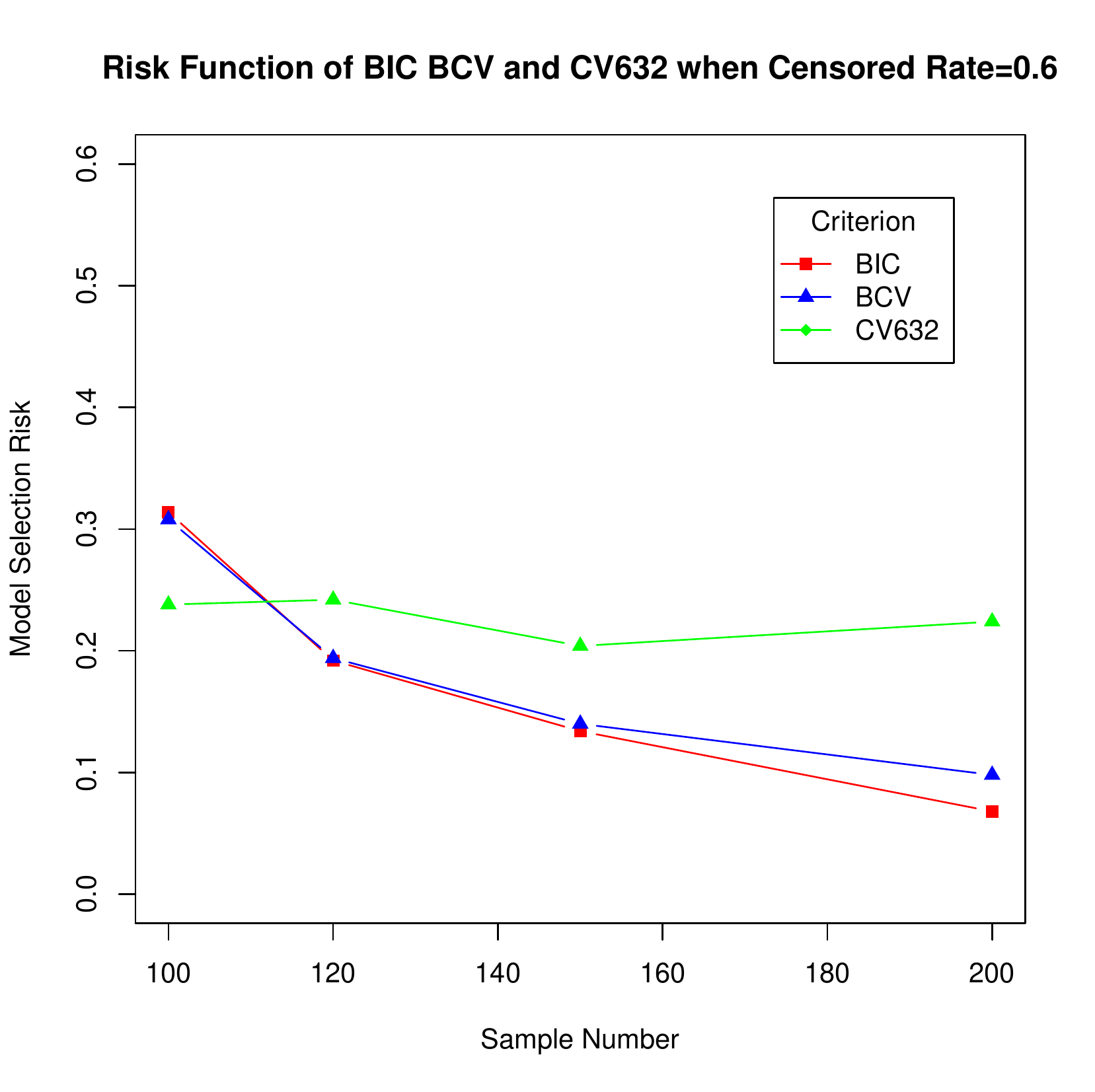}}
\subfigure[]{
\centering
\includegraphics[width=5.8cm]{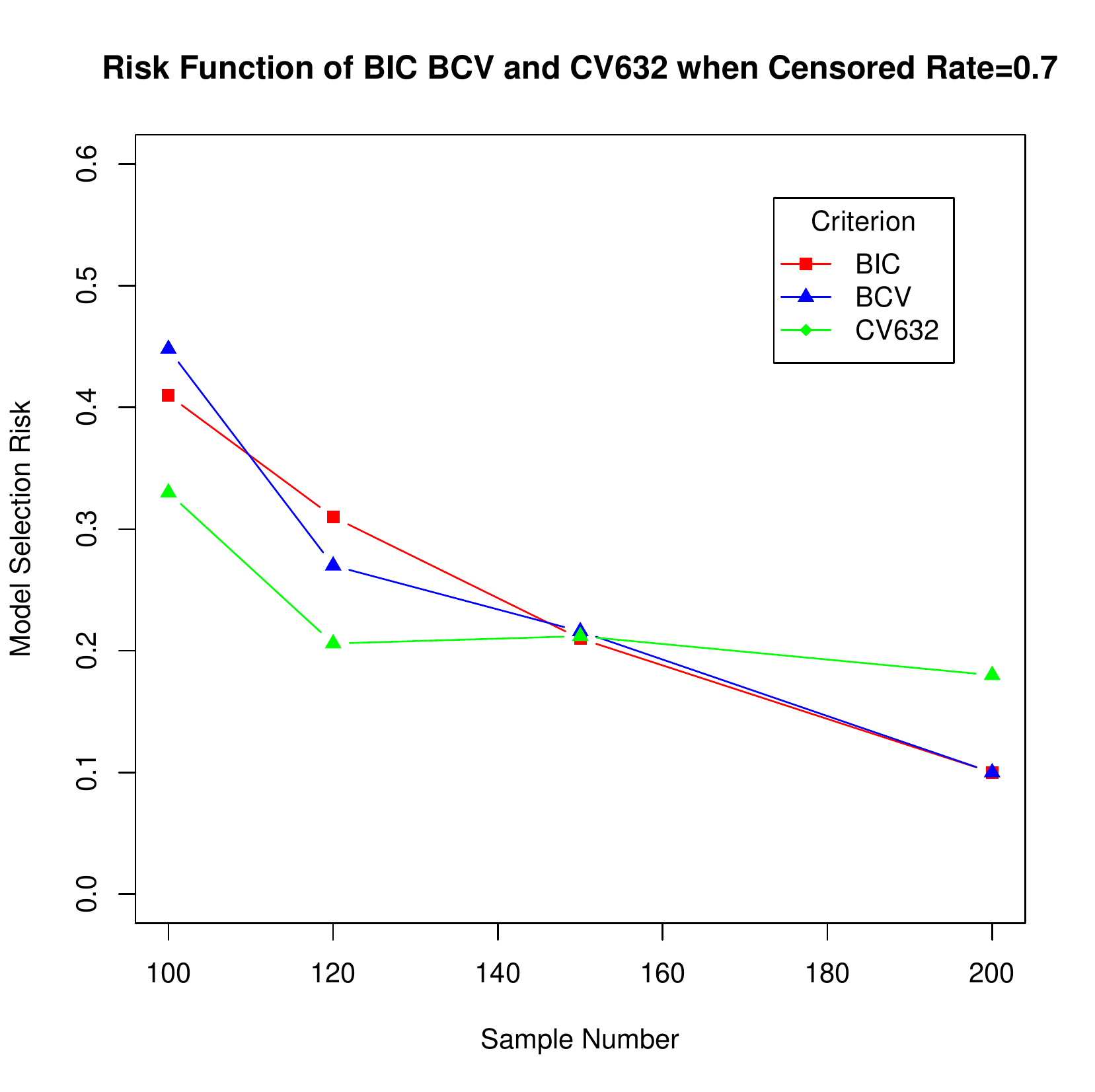}}
\subfigure[]{
\centering
\includegraphics[width=5.8cm]{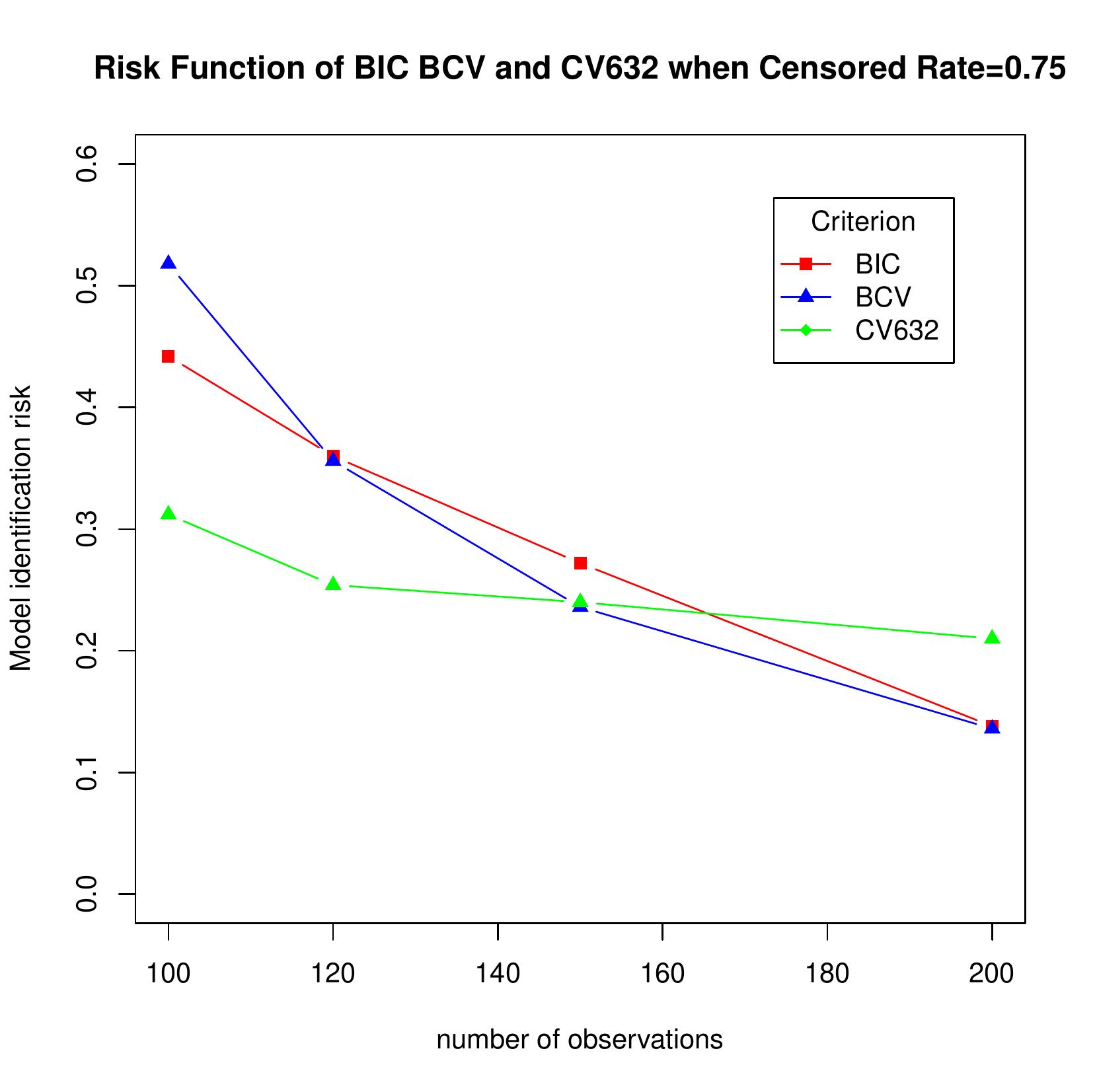}}
\end{figure}

The graph (a) and the graph (b) illustrate that the performance of the
BIC and BCV are superior to the CV632 criterion when the number of observations are taken as $n=200$ and $n=150$. However, the performance of the
CV632 criterion is becoming more competitive as depicted by the graph (c) and the graph (d). Specifically, as is shown by the graph (d), the performance of the CV632 criterion is becoming uniformly superior to the BIC and BCV model selection criteria when the number of observation is taken as $n=100$.

The graph (e) and the graph (f) demonstrate that the both the model identification risk of the BIC and BCV are decreasing with
the increase of the number of observations when the censoring rate are taken as fifty percent and sixty percent respectively.
Moreover, the performances of the BIC and BCV are superior to the CV632. However, the graph (g) and (h) show that the
performance of the CV632 is gradually becoming superior than the BIC and BCV criteria when the number of the observations varies from $n=100$ to $n=150$.
The intensive simulation experimentation demonstrates that the model selection performance of the $\rm CV632$ criterion becomes more competitive than its competitors
when the sample size is becoming limited and the censoring rate is becoming high.

\section{Real data analysis}

In this section, we apply the CV632 model selection criterion on the real data to demonstrate its model selection performance when the
Tobit regression model is used to fit real data under the circumstance of the inadequate observation information.

The real data set we use is the fidelity data, which is available at the package AER in the statistical analysis software R. The Affairs dataset contains a total
number of $601$ individuals which will be considered as the potential population.
The variable named Affairs is the interesting response variable with $0.75$ censoring rate and the rest $8$ variables will be taken as the potential possible
explanatory variables. There are totally number of $2^{8}$ possible candidate models and a total number of $C_{8}^{d}$ candidate models
for every specific candidate model family $\mathcal{F}(d+2)$, where $d, d=0,1,...,8$ is the number of explanatory variables got involved into the regression model.

In order to compare the model selection performance when the observation information is becoming inadequate, the model identification results
fulfilled by the BCV and BIC criteria based on the number of $601$ observation samples will be taken as the reference standard of the variable selection.
In order to demonstrate the performance of the model selection under the circumstance of the limited observation samples, a total number of $130$ observations
are sampled from the Affairs dataset.
The minimum values of the BCV and BIC criteria for any specific candidate model family $\mathcal{F}(k)$, $k=2,3,...,10$ are summarized in Table $5$.
The notations $BCV_{min}(d)$ and $BIC_{min}(d)$ stand for the minimum values of the BCV and BIC criteria of the specific candidate model family $\mathcal{F}(d+2), d=0,...,8$.
More specifically, the regression model family will reduce to
the simplest model family $\mathcal{F}(2)$ which only consists the intercept term $\beta_{0}$ and the disturbance term with
the number of explanatory variables $d=0$.
\begin{table}
\caption{The minimum BCV and BIC of the candidate model family $\mathcal{F}(k)$, $k=2,...,10$ based on the number of $601$
observation individuals}
\begin{center}
\setlength{\tabcolsep}{0.6mm}{
\begin{tabular}{cccccccccc}\toprule
\scriptsize
  $d $ & 0 & 1 & 2 & 3 & 4 & 5 & 6 & 7 & 8 \\ \hline
  BCV$_{min}(d)$  & 2097 & 1957 & 1439 & 1954 & \bf{1428} & 1431 & 1436& 1441 & 1445 \\
  BIC$_{min}(d)$  & 1502 & 1463 & 1457 & \bf{1449} & 1453 & 1456 & 1463& 1468 & 1474 \\
  \hline
\end{tabular}}
\end{center}
\end{table}

As is shown in Table $5$,
the estimation of the number of explanatory variables $\hat{d}=4$ for the BCV criterion and the corresponding variable combination is age, yearsmarried, religiousness and rating.
As for the BIC criterion, the estimated number of variables is $\hat{d}=3$ and the corresponding variable combination is yearsmarried, religiousness and rating.

The minimum BCV, BIC and CV632 of any specific candidate model family $\mathcal{F}(d+2)$, $d=0,1,...,8$ based on the number of $130$
observation individuals $BCV_{min}(d)$, $BIC_{min}(d)$ and $CV632_{min}(d)$ are summarized in Table $6$.

\begin{table}
\caption{The minimum BCV, BIC and CV632 of the candidate model family $\mathcal{F}(k)$, $k=2,...,10$ based on the number of $130$
observation individuals}
\begin{center}
\setlength{\tabcolsep}{0.6mm}{
\begin{tabular}{ccccccccccc}\toprule
  \scriptsize
  $d$  & 0 & 1 & 2 & 3 & 4 & 5 & 6 & 7 & 8 \\\hline
  BCV$_{min}(d)$ &468  &391  &\bf{291}  &412  &296  &302  &307 &314  &322  \\
  BIC$_{min}(d)$  &307  &\bf{306}  &307 &307&311  &316  & 321 &327 &343   \\
  CV632$_{min}(d)$  & 405 & 356 & \bf{290} & 367 & 293 & 295 & 298& 302 & 312 \\
  \hline
\end{tabular}}
\end{center}
\end{table}

Table $6$ shows that the estimated number of explanatory variables for CV632 criterion is $\hat{d}=2$ and the corresponding variable combination is age and yearsmarried.
As for the BCV criterion, the final estimated number of explanatory variables is also $\hat{d}=2$ with $BCV_{min}(2)=291$; however, the corresponding variable combination is age and children. The estimated number of variables for the BIC criterion is $\hat{d}=1$ with $BIC_{min}(1)=306$ and there is only one explanatory variable named rating got involved into the model.

The real data analysis results demonstrate that the performance of CV632 will be superior to the performance of both the BCV and BIC criteria when
the observation information is becoming limited. The performance of the real data analysis of the CV632 criterion is consistent with
its performance in the simulation study section.

\bibliographystyle{unsrt}
\bibliography{mybib}

\end{document}